# Protection of Accelerator Hardware: RF systems

*S.-H. Kim*
Oak Ridge National Laboratory, Oak Ridge, TN, USA

**Abstract**
The radio-frequency (RF) system is the key element that generates electric fields for beam acceleration. To keep the system reliable, a highly sophisticated protection scheme is required, which also should be designed to ensure a good balance between beam availability and machine safety. Since RF systems are complex, incorporating high-voltage and high-power equipment, a good portion of machine downtime typically comes from RF systems. Equipment and component damage in RF systems results in long and expensive repairs. Protection of RF system hardware is one of the oldest machine protection concepts, dealing with the protection of individual high-power RF equipment from breakdowns. As beam power increases in modern accelerators, the protection of accelerating structures from beam-induced faults also becomes a critical aspect of protection schemes. In this article, an overview of the RF system is given, and selected topics of failure mechanisms and examples of protection requirements are introduced.

## 1 Introduction

Particle acceleration using time-varying electromagnetic fields is called radio-frequency (RF) acceleration. Radio-frequency system frequencies for particle accelerators range from 10 MHz to 30 GHz, either in continuous-wave or pulsed operation. The RF power per unit station ranges up to a few megawatts in continuous-wave machines and 100 MW in pulsed machines. Many state-of-the-art technologies are involved in producing the RF systems, such as vacuum science, high-voltage technology, surface physics, advanced materials, and high speed controls. As superconducting RF technologies become a choice for modern machines, cryogenics, superconducting RF science, and ultra-clean processing play important roles for RF systems.

A typical layout of an RF system for a particle accelerator is shown in Fig 1. The high-power chain is depicted with solid lines and the low-power part with dashed lines. For proton or ion acceleration, various types of accelerating structure are needed because each type of accelerating structure has a limited acceptance of particle velocity. Figure 2 shows different types of accelerating structure using normal conducting and superconducting technologies. As the beam power or energy stored in the beam increases in modern accelerators, the protection of accelerating structures from the beam becomes more important in machine protection schemes. Also, the accelerating gradients required for future machines are increasing, and there are questions over the breakdown limits, in terms of accelerating gradient, pulse length, trip rate, and expected lifetime in connection with material type, processing method, and processing history.

The goal of hardware protection is to keep the system in a reliably operable condition until the end of the equipment's lifetime. The concepts for this would be classified in two categories; setting the normal operating parameters to ensure their lifetime and minimizing abnormal operating conditions that could damage the system. There are many mechanisms that could cause damage and lead to catastrophic failures, e.g.:

– water leaks into the system components;
– air leaks into the vacuum boundary spaces;
– sharp edges in high-electric-field regions;
– dirty RF surfaces;
– resonant electron multipacting;
– condensed or trapped gas on RF surfaces;
– large reflected power to RF sources;
– over-power (voltage or current) to system components;
– beam bombardments on RF surfaces: activation, errant beam, mis-steered beam.

The specific requirements for the system protection vary depending on equipment-specific characteristics, such as machine type, beam power, beam energy, or beam pulse length. Since the hardware protection system mostly deals with abnormal conditions, it is essential to have a good understanding of the physics of individual equipment or components, and their interplay with the system. This understanding is important not only for normal operating conditions but also for any possible upset conditions.

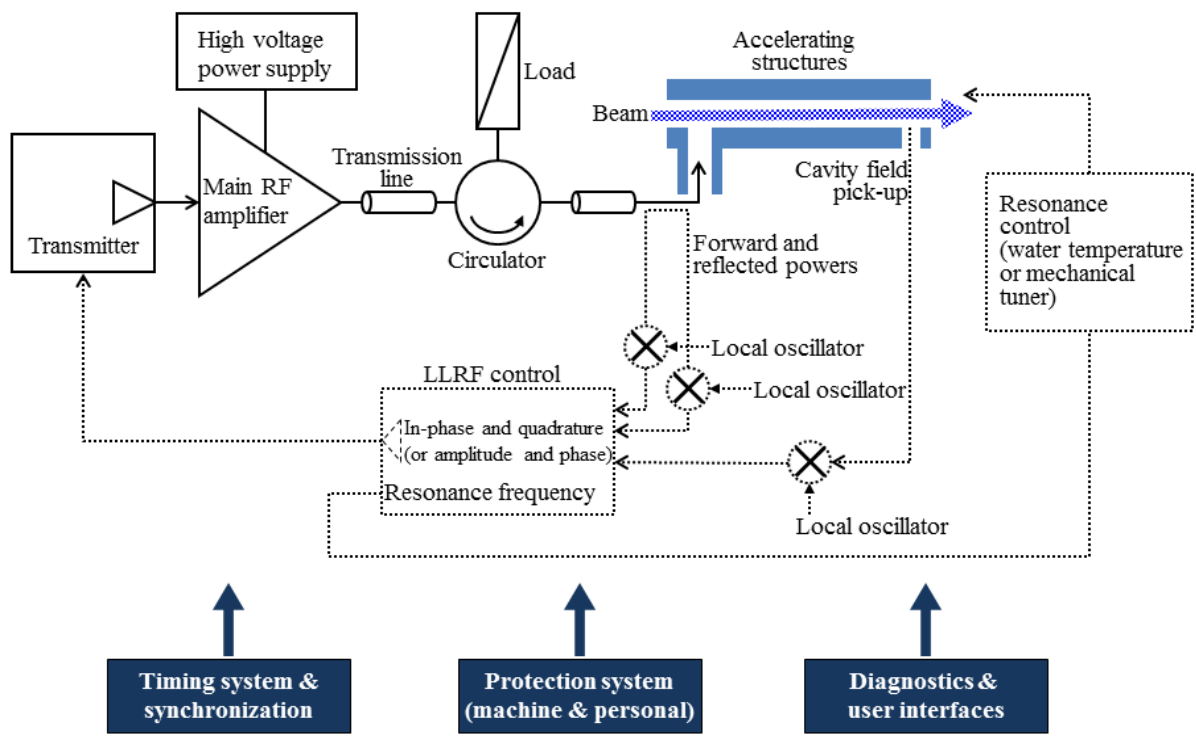

Fig. 1: Typical RF system for typical particle accelerators

One of the most threatening consequences in RF systems is the generation of surface damage, which could be caused by several different mechanisms. For example, arcing or discharge could occur in a system; interaction with large quantities of available RF power or stored energy in the system could result in severe arcing conditions and consequent surface damage. Another example might be a mis-steered beam that directly strikes the accelerating structure and could also cause surface damage. If an event is mild, a system will be brought to a normal condition through conditioning processes. However, when surface damage happens at a location of high electric field, ceramic surfaces, welding or brazing joints, or aggressive multipacting regions, the process could be irreversible. The severity of arcing events is determined by local physical conditions. In many cases, readings from diagnostic devices, such as vacuum gauges, beam loss monitors, or temperature sensors, do not reveal actual local conditions. Without a thorough understanding of the failure mechanism, it may not be possible to detect a precursor of the threatening events. When systems are over-protected, beam availability will be reduced. Thus, the protection system should be designed to avoid or minimize these threatening events, while maintaining beam availability.

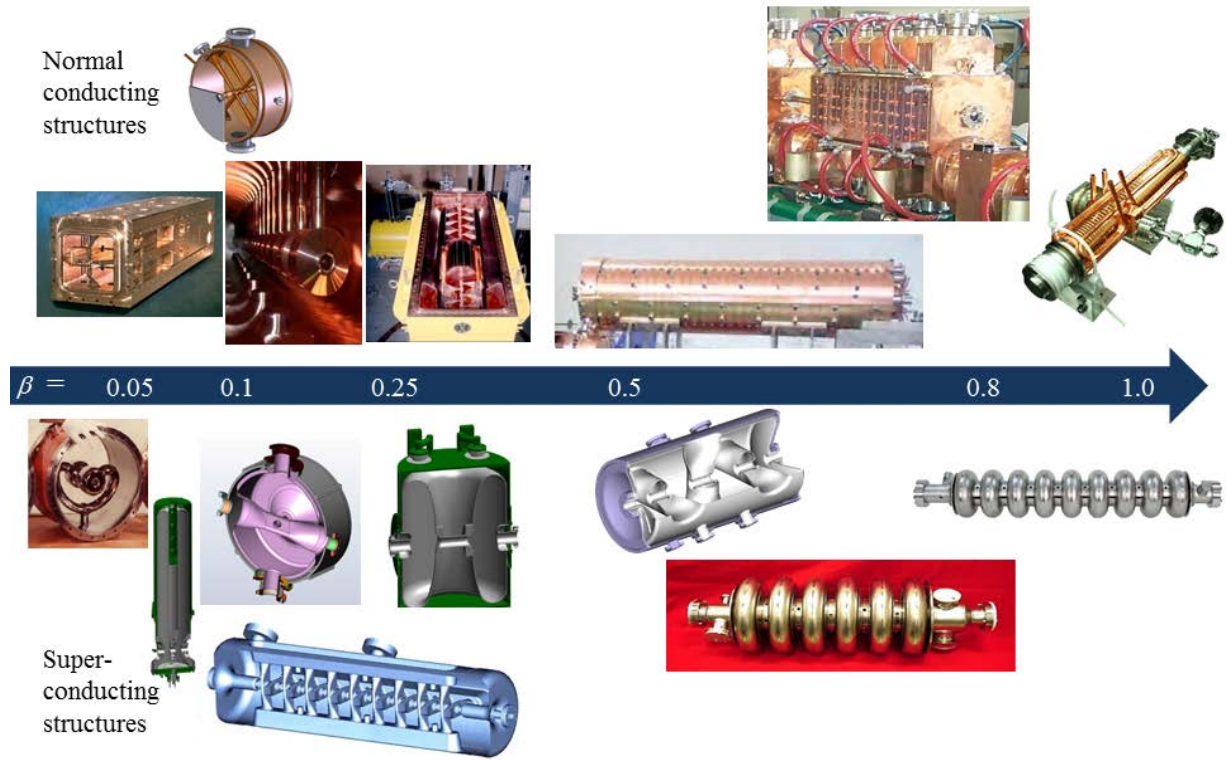

**Fig. 2:** Accelerating structures for particle acceleration

## 2 Breakdown

### 2.1 Vacuum breakdown mechanism

There has been a great deal of work to understand mechanisms for the vacuum breakdown, e.g. in high-voltage, plasma, particle accelerator, or vacuum science. In a high vacuum, the electron mean free path is much longer than the distance between electrodes or the length of RF structures, which means that there is no formation of electron avalanches in space for the initiation of breakdown, as in gas glow discharge. The details of physics in vacuum breakdowns are not yet well understood; however, several models developed by various authors are available. There are differences between RF and d.c. breakdowns. In an RF field, electrodes (cathode and anode) are not defined, as in d.c. fields. One can consider surfaces where electrons are emitted as cathodes and surfaces hit by electrons as anodes in RF breakdown. In this context, cathode and anode surfaces in an RF field reverse every half cycle. There is also a difference related with the duty factor. However, RF and d.c. breakdown have similar underlying physics. Studies to explain vacuum breakdown are focused on mechanisms that trigger a breakdown, such as the release of electrons and gases into the vacuum space. Thus, breakdown models can be classified in terms of breakdown initiation. Selected mechanisms are summarized in the following sub-sections.

## 2.1.1 Field emission model

As the electric field at a surface is increased, the potential barrier at the surface that bounds electrons inside the material becomes narrower. At a very high field, this potential barrier is so narrow that some electrons can pass through it. This effect is known as the tunnelling effect and results from the quantum mechanical nature of electrons. The resulting electron emission is called field emission, and is also known as cold emission. The Fowler–Nordheim equation predicts the current density from field emission,

$$j \propto \frac{(\beta E_s)^2}{\varphi} \exp\left[\frac{-a\, \varphi^{3/2}}{\beta E_s}\right],$$

where $\beta$ is the field-enhanced factor determined by the emitter geometry, $E_S$ is the surface electric field, $\varphi$ is the work function of the surface, and $a$ is the constant.

Peak surface electric fields of accelerating structures are normally higher than 10 MV/m and up to a few hundred megavolts per metre. X-ray measurements show the existence of electrons and their accelerations in normal operating conditions. This current is often called the dark current and this effect is one of electron loading in the RF systems. In superconducting RF cavities especially, it could limit the achievable accelerating gradient, reduce the quality factor, and quench local spots or the whole system. If field emission passes a threshold, breakdown happens, with sharp increases in the emissions of X-rays and electron currents. Figure 3 shows images on a phosphor screen installed at the end of superconducting RF cavities during testing. The bright spots in the image indicate X-rays and electron bombardments from field emission. As long as the field-emitted electron currents remain in pre-breakdown condition, the system is stably operable. As the dark current increases, and if the acceleration of field-emitted electrons is efficient, the vacuum pressure could be increased, and eventually affect operational stability.

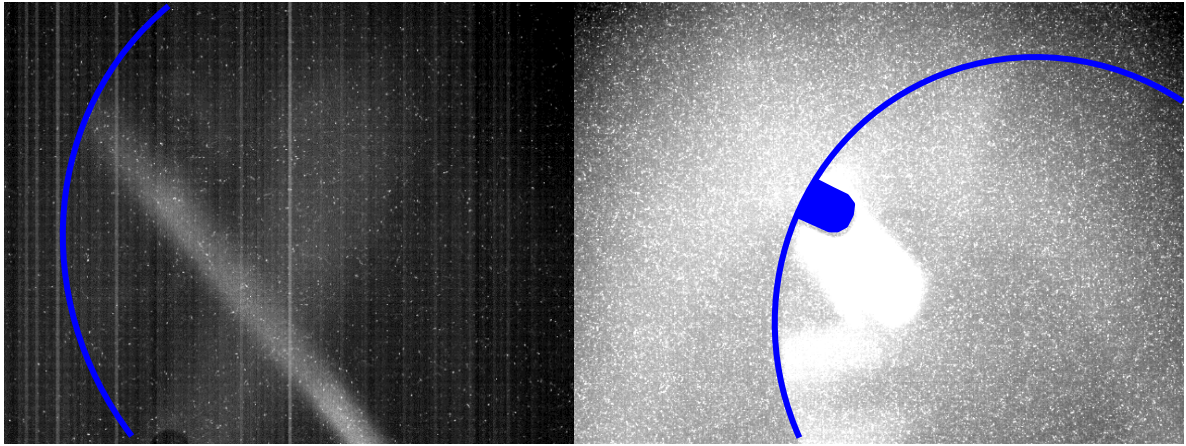

**Fig. 3:** Phosphor screen images of X-rays and electrons from field emission

The theoretical field emission threshold is about 1 GV/m; however, field emission is observed even as low as 10 MV/m in many RF systems, presumably owing to other enhancing factors. Models for the field emission enhancement are:

– protrusion-to-protrusion;

– absorbed gases and other contaminants;

– activation of field emitter at elevated temperature;

– dielectric layer.

It is quite difficult to quantify the characteristics of field emitters, owing to their complex nature, with factors such as size, shape, species, charge status, binding status on the bulk material, temperature, or process history to be taken into account. The emitters are statistically distributed over surfaces and the control of emitters during fabrication, processing, and operation is quite challenging. Figure 4 shows examples of field emitters observed on samples in a study of particulate control during the surface process for superconducting RF cavities.

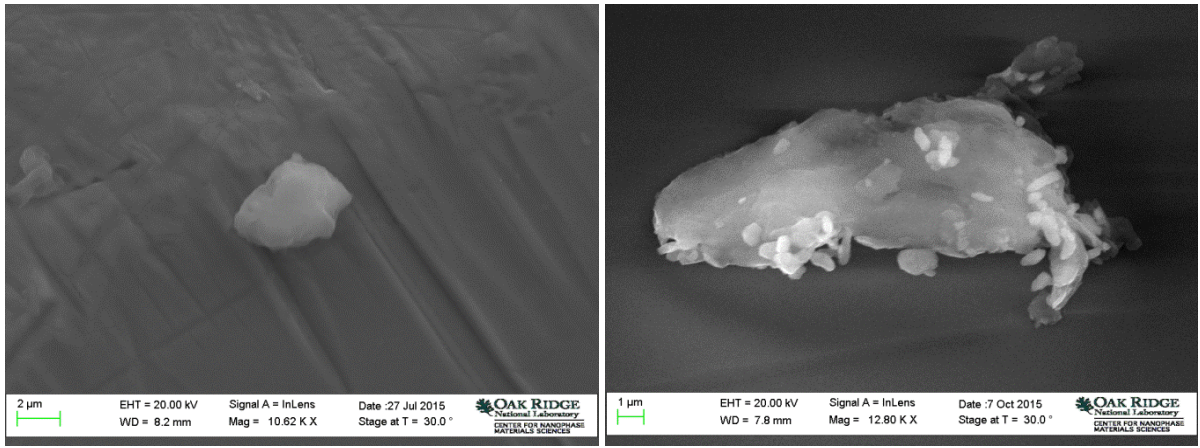

**Fig. 4:** Scanning electron microscopy images of potential field emitters observed on accelerator material samples (EHT, acceleration voltage; Signal A, electron detector mode; WD, working distance; Mag, magnification).

The field emission model for the initiation of vacuum breakdown assumes the existence of field emitters on surfaces. One mechanism for vacuum breakdown by field emission is that field-emitted electrons hit other surfaces, or anodes, causing local temperature increases and resulting in gas desorption. The desorbed gases can be ionized by the primary electrons. The ions can then increase the emission of primary electrons, owing to space charge formation, or generate secondary electrons by hitting surfaces. The process can continue and eventually lead to vacuum breakdown, in a condition similar to Townsend gas discharge. This model of d.c. breakdown is called the anode heating mechanism, and describes a breakdown mechanism with an avalanche of ionization around the anode. The other mechanism of vacuum breakdown by field emission is called the cathode heating mechanism. Field emission provides a pre-breakdown current at the field emitter. This current could heat up the emitter region, by resistive heating. When the emitter reaches a critical condition, it will melt and vaporize. This heating can increase the field emission current density. Under this condition, very dense local plasma is formed and craters, called cathode spots, are produced [1].

### 2.1.2 *Multipacting model*

Another major electron-loading effect in RF systems can come from multiple impacts of electrons, called multipacting. Multipacting can occur when both of the following conditions are met: (1) electron motions meet a resonant condition through interaction with an electromagnetic field in vacuum and (2) the secondary electron emission process develops more than one electron per one electron impact.

Secondary emission is a determining parameter for multipacting conditions and is characterized by the secondary emission yield, which is, the statistically measured number of emitted electrons per an incident electron. The secondary emission yield depends not only on material type but also surface treatment history, such as baking, gas discharge cleaning, coating, and impurity contents [2]. Figure 5 illustrates how the secondary emission yield changes depending on surface cleaning or treatment.

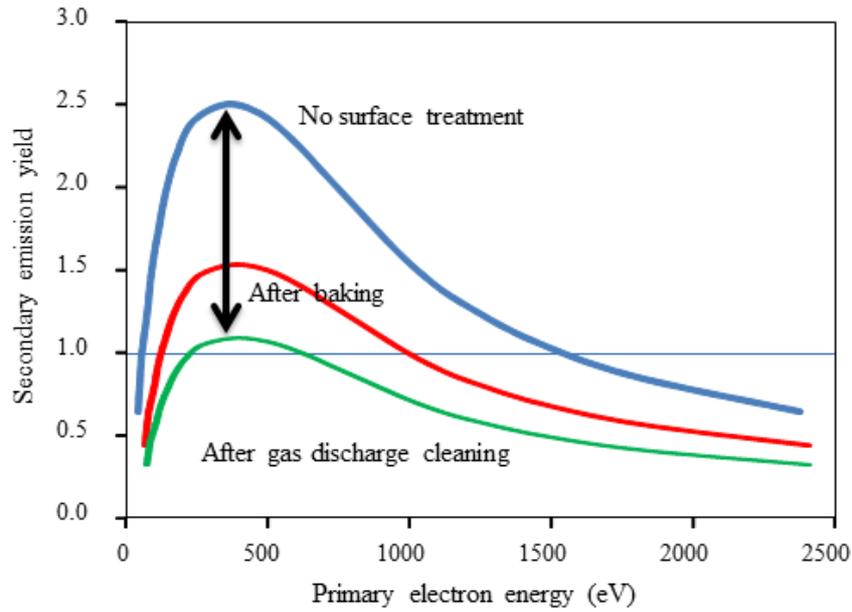

**Fig. 5:** Illustration of secondary emission yield changes

The resonance condition for multipacting is determined by the RF field profiles and field level. Figure 6 shows examples of electron trajectories under interactions with electromagnetic fields. If the time interval between impacts is $nT$ or $(n-1/2)T$ (where $n = 1, 2, 3\ldots$ and $T$ is the RF period), depending on multipacting type, for so-called one point and two point multipacting, emitted electrons will have similar starting conditions to the primary electron condition. Under this circumstance, and when the secondary emission yield is larger than unity, an electron avalanche will happen. Because of this condition, the multipacting barrier typically has bands. As indicated from the secondary emission yield, high-energy electrons do not play a role in the multipacting barrier.

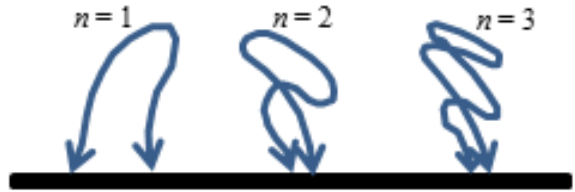

**Fig. 6:** Electron trajectories in resonance conditions

Multipacting usually happens at low-electric-field regions in RF structures. Low-electric-field regions can be in RF windows, irises, couplers, beam pipes, or equators of cavities and are prone to multipacting barriers. When an acceleration gap is much shorter than the wavelength, multipacting can happen around the acceleration gap, where the electric field is high. In pulsed machines, multipacting can occur during the RF ramp-up and decay periods of each pulse, if the multipacting barrier exists below an operating field. During the process of multipacting, the non-resonant portion of electrons will escape the multipacting trajectory and can be accelerated in accelerating structures. These electrons heat the surfaces and can generate X-rays, in the same way as electrons from field emission, and give rise to thermal instability, especially in superconducting RF cavities. Figure 7 shows an example of X-rays from multipacting. From this plot it is not clear whether the X-rays result from field emission or multipacting. By monitoring changes of radiation waveform in pulsed operation as a function of field level, multipacting can be verified relatively easily.

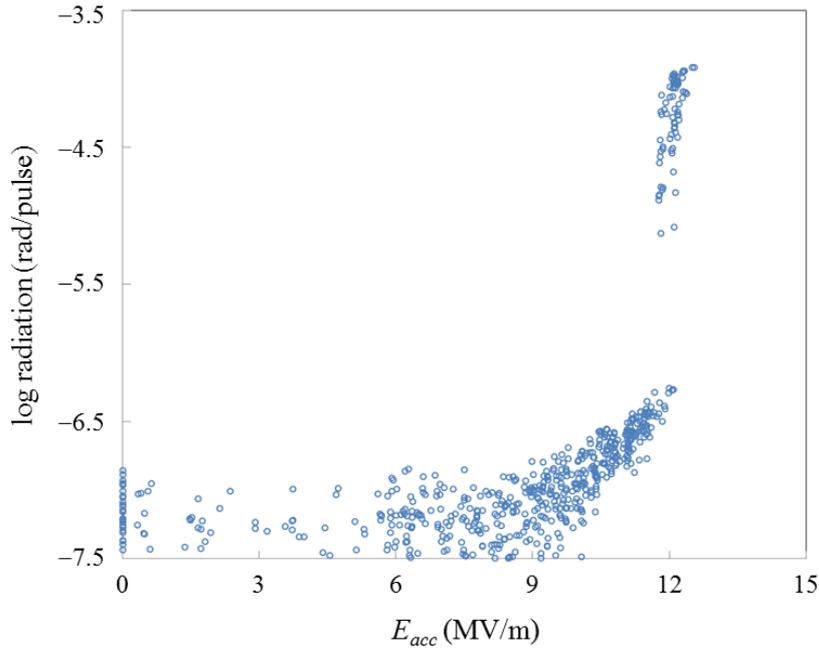

**Fig. 7:** X-rays by non-resonant electrons from multipacting

In many cases, the multipacting barrier can be processed out by careful RF conditioning. However, if severe multipacting develops, the multipacting region will heat up, and trapped or condensed gases will be released from the surface. This condition can eventually cause vacuum breakdown and a material defect can occur. This kind of breakdown has been observed in almost every RF systems in vacuum, such as electron tubes, couplers, waveguides, RF windows, and accelerating structures. To avoid a severe multipacting barrier, careful analysis should be done from the design stage. As mentioned earlier, surface processing plays an important role; one must keep the surface clean and avoid contaminants, perform careful RF conditioning, bake surfaces, apply discharge cleaning where possible or surface coatings with low-secondary-emission-yield material, and apply d.c. biasing to shift multipacting barriers.

### 2.1.3  Particle exchange model

This model assumes that a charged particle comes out of a surface, which is always possible statistically, then hits other surfaces of the electrode and liberates particles. Oppositely charged particles are accelerated back to the initial surface of the electrode. If this process becomes cumulative, it could trigger vacuum breakdown. This mechanism is based on avalanching of mutual secondary emission of ions and electrons, and also photons and absorbed gases on surfaces.

### 2.1.4  Clump model

This model assumes that a loosely bound particle cluster is on the surface. In a high voltage, the cluster is then charged and accelerated and impacts other surfaces, releasing gases and vapours, and triggers vacuum breakdown.

### 2.1.5  Kilpatrick criterion

In 1950s, W. Kilpatrick developed the criterion on RF breakdown from experimental data [3]. The Kilpatrick empirical formula relates the breakdown electric field with RF frequency;

$$f = 1.64 \, E_s^2 \exp\left(-\frac{8.5}{E_s}\right),$$

where $f$ is the RF frequency in MHz and $E_s$ is the breakdown electric field at the surface in MV/m. The experimental data for this criterion were generated when clean vacuum systems were not available. Modern accelerating structures are designed for and run at higher fields than this criterion; however, it is still in use as a kind of figure of merit. It is interesting to note that some much later analyses using anode heat mechanism [4] and experimental data [5] show similar dependencies on RF frequency. However, there is no good explanation for this similarity. Conversely, it has been reported [6] that no increase of breakdown field is observed at frequencies higher than 10 GHz.

## 2.2 Dielectric breakdown

Dielectric materials are widely used in various components of RF systems to separate electrodes or form a boundary between vacuum and air. Failures of dielectric materials in RF systems are among the most frequent and severest events. Figure 8 shows examples of dielectric material failures. Dielectric materials have intrinsic breakdown limits, and actual operating conditions are typically set well below these intrinsic limits. However, electrical breakdown also occurs well below the intrinsic limit. Possible enhancing factors for dielectric breakdown would be charged-particle bombardment or non-uniformity of the dielectric material. These enhancing factors provide initiation of breakdown, generate field concentration, and can lead to non-uniform charge build-up. Sources of non-uniformity are voids, impurities, inhomogeneity of the material, non-ideal boundary junctions, absorbed gases, and wrinkles. A selection of mechanisms practically important for RF systems is briefly reviewed next.

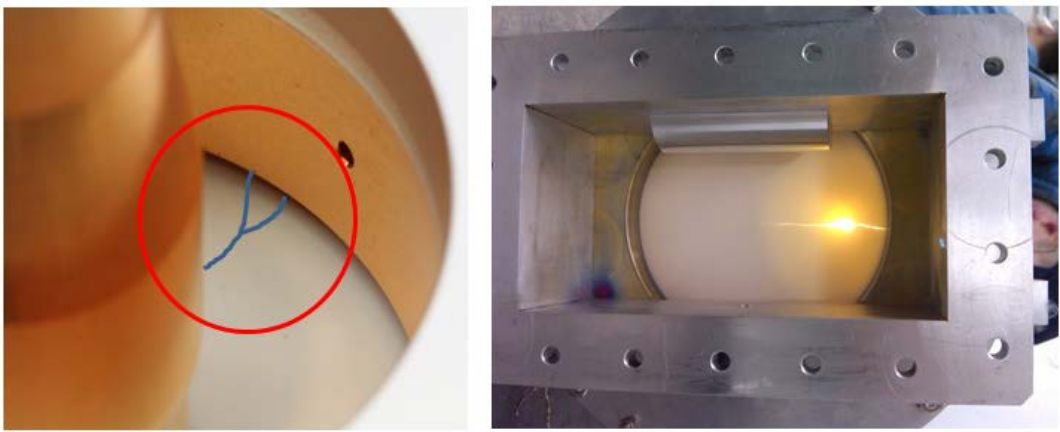

**Fig. 8:** Left: 550 kW, 805 MHz coaxial power coupler RF window crack. Right: 5 MW, 805 MHz RF window failure.

- Chemical deterioration: Dielectric materials facing air or other gases could undergo oxidation or hydrolysis processes that give rise to surface cracks or changes in material properties. Operation at an elevated temperature would accelerate these unwanted chemical processes. Even dielectric materials under vacuum could have chemical deterioration through particle bombardments. For example, an electronic activity such as multipacting on an alumina surface results in oxygen defects and leads to severe breakdown [7].

- Treeing: When a dielectric fails, sometimes a visible conducting path or pinhole across the dielectric is observed. The leakage current passes through the conducting path, leading to sparks. The spread of spark channels is called 'treeing', since it looks like the branches of a tree. Treeing is a pre-breakdown phenomenon and a damaging process from partial discharges. At the edge of a treeing there is a high field concentration. When a system runs for a long time under electrical stress, the treeing progresses through the dielectric, finally leading to failure. Initiation of this process comes from inhomogeneity of the surface, contaminants on the surface, or charge build-up from other sources.

- Breakdown due to voids: Dielectric materials have voids within the medium or at junction boundaries. Voids are usually under vacuum or filled with gases whose electric permittivity is smaller than that of the dielectric. The electric field strength is, therefore, higher in the voids. When the field strength in the voids is higher than the voltage hold-off capability, local breakdown occurs. In this case, the effect of local discharges and their developments in the voids is similar to treeing.
- Surface flashover: Surface flashover typically occurs at a much lower electric strength than volume breakdown. Experimental measurements show that the surface of a dielectric in a vacuum becomes positively charged under a high-voltage d.c. by electron multiplication through a secondary emission process [8] or an electron cascade in a thin surface layer [9]. Once electron multiplication is developed, travelling electrons form a pre-breakdown current and release gases from the surface by electron-stimulated desorption. The desorbed gases are ionized by electron collisions and breakdown occurs along the surface. It is broadly agreed that the initiation of flashover starts at a so-called triple junction point (Fig. 9). In practice, voids or small gaps exist at the joint, where the electric field strength is much higher than the majority of the bulk area. Similar phenomena could be initiated with charged-particle bombardments that are produced somewhere else.
- Mechanical failure due to thermal shock: An arcing event generates a thermal gradient in a short period of time that results in mechanical stress due to a localized thermal expansion. If the event is large enough, the mechanical stress can reach a rupture condition.

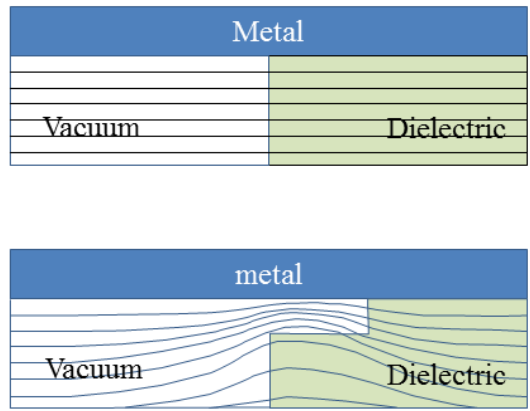

**Fig. 9:** Electric field profile at the triple junction (top, ideal case; bottom, practical case)

## 2.3 Vacuum

As discussed earlier, in a vacuum condition, released gases from the surface play a particularly important role in breakdown. Mechanisms for gas release can be classified as follows.

- Desorption is the process of gas release from the surface and is the final state of all outgassing mechanisms. The binding energy of adsorbed gases is much less than an electron volt. Thus, the desorption rate increases at an elevated temperature and a much higher desorption rate can be achieved with energetic charged-particle and photon bombardments.
- Vaporization is the phase transition of a material into a gas. Materials with higher vapour pressure can evaporate in a vacuum or at an elevated temperature.
- Dissolved gases in bulk materials, such as metals and ceramics, move to the vacuum surface. This process is called diffusion. Hydrogen has a high mobility in the bulk material.
- Adsorbed gases from the outside surface diffuse through the bulk material and are then desorbed from the surface in a vacuum. This process is called permeation.

Reducing outgassing is the critical process for stable operation of electron tubes, accelerating structures, power couplers, and high-power RF windows. Polishing reduces the effective surface area and capacity of adsorption. Baking and vacuum firing are the most widely used processes for reducing outgassing [10, 11]. Baking in vacuum at moderate temperatures (150°C–300°C) reduces the outgassing rate, especially for water. This moderate baking in vacuum is not sufficient to remove hydrogen from the bulk material. High temperature vacuum firing, however, drastically reduces the hydrogen outgassing rate. During the baking or vacuum firing, $H_2$, CO, $CO_2$, and $CH_4$ gases are usually released along with water. To further clean the surface, gas discharge cleaning or electron beam showering is used.

## 3 Beam related issues

One of the most important purposes of hardware protection is to reduce equipment activation levels to as low as reasonably achievable, to allow hands-on maintenance and protect machine equipment from damage due to uncontrolled beam strikes. Many groups have studied allowable uncontrolled beam loss and it is generally agreed that an average beam loss of 1 W/m is a reasonable limit for hands-on maintenance [12, 13]. As demands for beam power and beam power density increase beyond previously experienced levels, the operation of high-intensity machines below this limit will be very challenging. Currently, there is a great deal of effort with computer simulation, wide dynamic range diagnostics, and fast beam abort schemes in response to upset conditions, to reduce uncontrolled beam losses. Since other sessions in this course cover details of beam dynamics and equipment activation issues, in the following other possible sources of beam-induced damage to accelerating structures will be introduced.

### 3.1 Mis-steered beam

When a beam is steered by mistake or as a result of a magnet fault, it can strike the surface of the accelerating structure. The stopping power, which is the loss of particle energy per unit path length, for electrons is rather uniform, while that for ions has a peak before stopping, known as the Bragg peak. Figure 10 shows trajectories and the stopping power of electrons in a simulation of the Spallation Neutron Source (SNS) linac beam dump window. The electron energy in this example is ~1 MeV, stripped from the $H^-$ ion. Figure 11 shows an example of Bragg curves for 2.5 and 7.5 MeV protons.

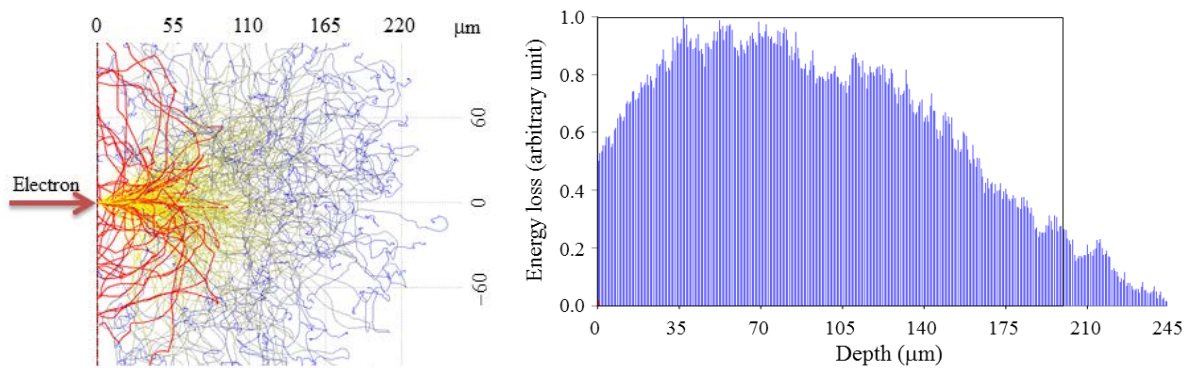

**Fig. 10:** Example of (left) electron trajectories and (right) energy loss per unit length in nickel alloy

When particles lose their energy while passing through matter, a good portion of energy will be converted into kinetic energy; hence, the local temperature will increase. It is important to understand the dependencies of beam parameters on damage to accelerator structures. Thus, the machine protection system for aborting the beam should be designed accordingly. In the following analysis, a guide to determine the beam abort speed is discussed, using the SNS beam parameters.

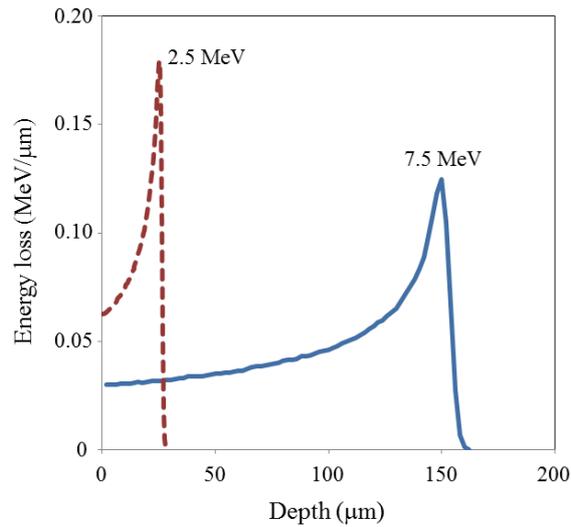

**Fig. 11:** Example of energy loss per unit length of proton in copper

Figure 12 shows an example of the thermal load from the round Gaussian beam with $\sigma = 4$ mm at 20 mA, with 7.5 MeV protons. Since the heating effect by the beam is localized, a very steep thermal gradient will occur and result in thermal stress. If the stress level exceeds a mechanical stress limit, such as the yield point or tensile strength, mechanical defects could occur. Usually, this mechanical stress threshold comes much earlier than material melting. Figure 13 shows the peak stress development in copper as a function of beam strike duration. In this example, the peak temperature is around 200°C when the peak von Mises stress from the thermal gradient reaches the tensile limit.

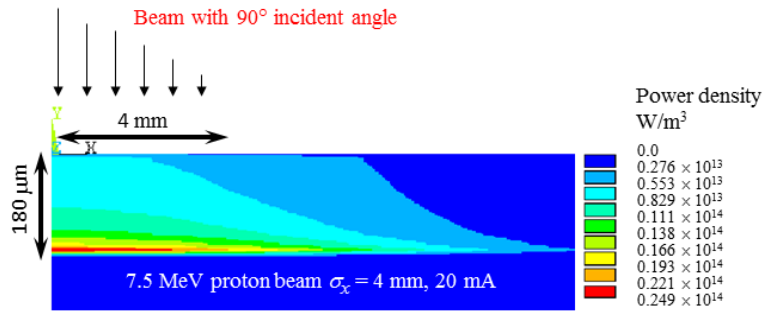

**Fig. 12:** Thermal load profile in copper from beam strikes

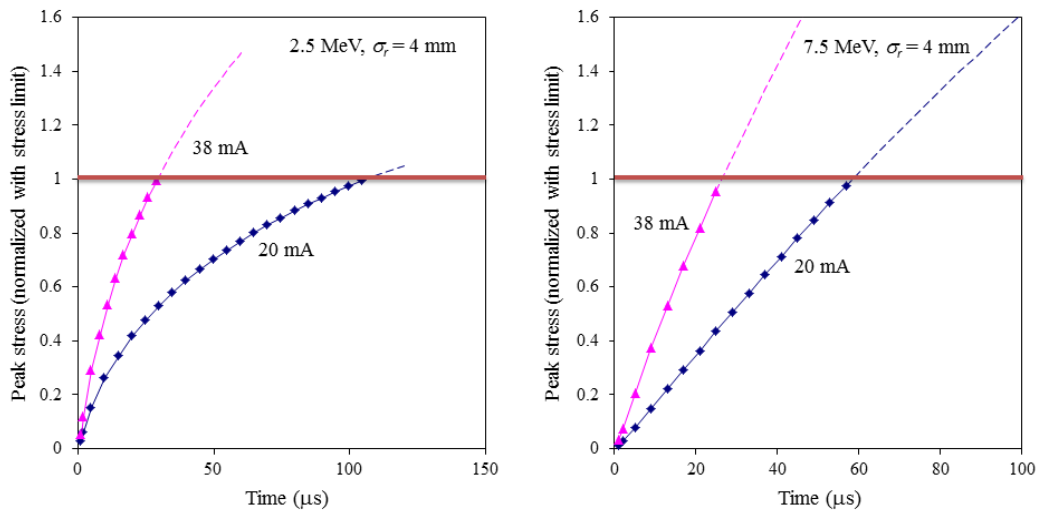

**Fig. 13:** Stress development from different beam conditions at 90° incident angle

Using the same model, one can explore stress limits under various beam conditions, as shown in Fig. 14. In practice, when a mis-steered beam condition occurs, the beam is most likely to strike at a grazing angle on the accelerating structure. Figure 15 shows an example of the energy loss per unit length at various incident angles. The energy loss per unit length at grazing angles increases as the incident angle reduces but the effective beam size also increases rapidly. The depth is measured along the direction normal to the surface. Beam dynamics simulations predict that the worst probable case for the SNS accelerator not more than 3°. Figure 16 shows an example of the allowable duration for a mis-steered beam for grazing angle incidents compared with the 90°-incident cases.

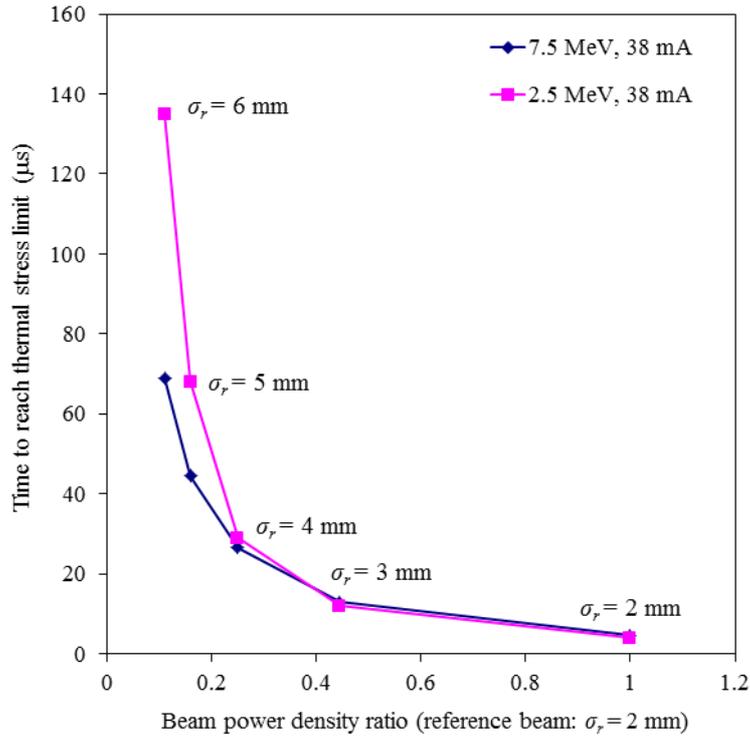

**Fig. 14:** Time to reach the mechanical stress limits for various beam condition at 90° incidence. In this example, a beam with $\sigma_r$ = 2 mm is used as a reference nominal beam size.

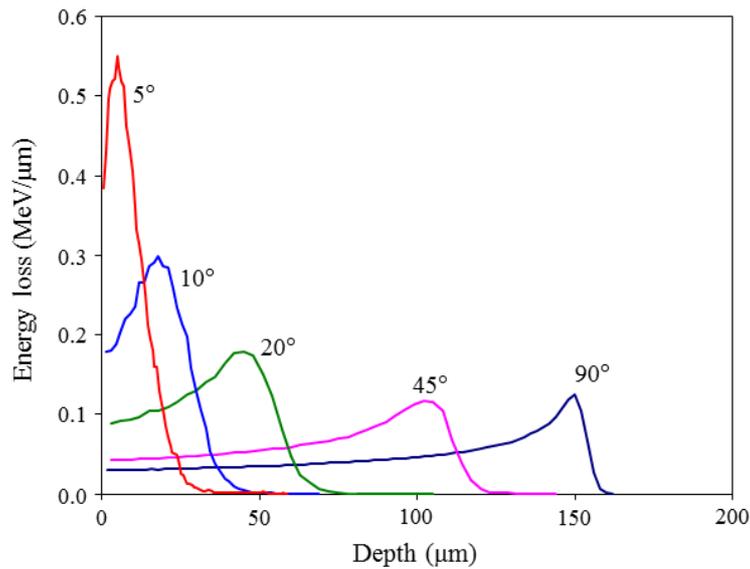

**Fig. 15:** Energy loss per unit length for various beam incident angle for protons with 7.5 MeV

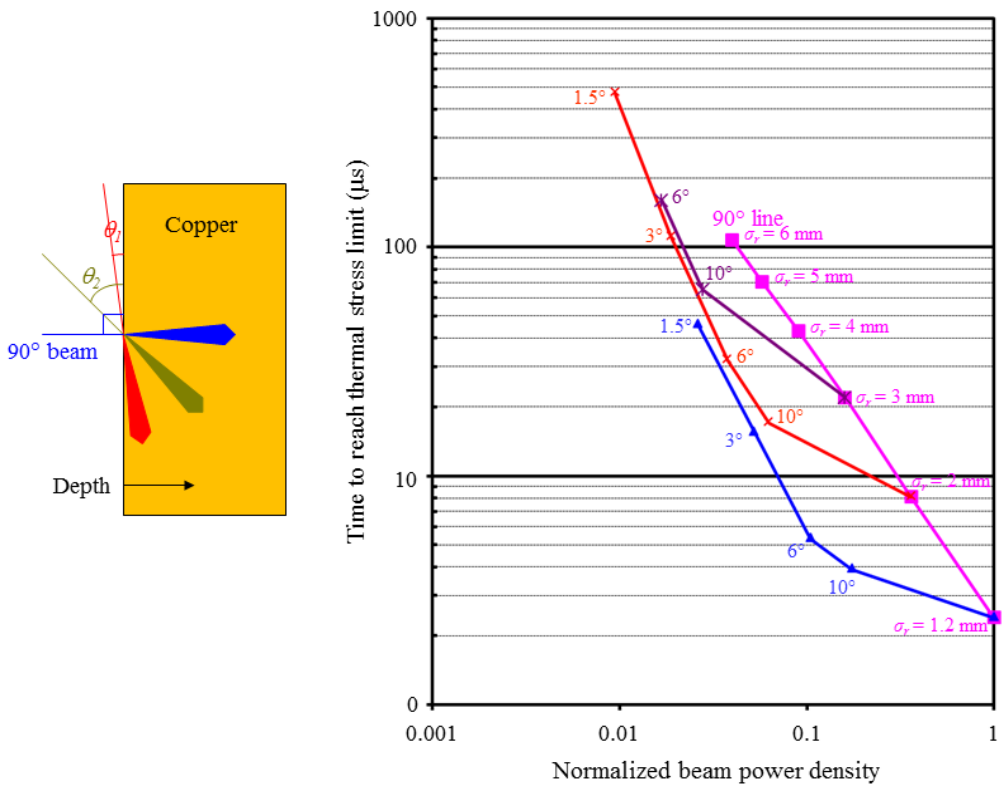

**Fig. 16:** Time to reach mechanical stress limits from 7.5 MeV, 26 mA beam at various beam sizes and incident angles.

## 3.2 Errant beam

The term 'errant beam' hereafter refers to an off-energy beam generated anywhere in the accelerator, and transported downstream in a fault condition; this is different from the uncontrolled beam loss in a normal operating condition. Since the errant beam is an off-energy beam, it is mostly lost while transported through the linac, resulting in beam trips in the linac. During an errant beam condition at the SNS linac, the beam loss region ranges from several to ten cryomodule lengths or longer. Two mechanisms are identified as sources of errant beams at the SNS: (1) RF truncations in upstream RF structures and (2) abnormal beam generation in the front end, which includes the ion source, the low energy beam transport system, and the RF quadrupole.

The SNS linac (Fig. 17) is composed of the front-end system, 6 drift tube linacs, 4 coupled cavity linacs, and a superconducting linac that houses 81 superconducting RF cavities in 23 cyromodules. Each RF structure has its own fast interlocks, which truncate an RF pulse at an upset condition, such as a vacuum burst, discharge, arc, or any RF signal beyond a predefined threshold. These interlocks trigger the machine protection system, which aborts the beam. There is a delay time for the machine protection system to abort the beam, as illustrated in Fig. 18. When an interlock of any RF structure truncates its RF pulse at an upset condition, the beam from the start of the RF truncation up to beam abortion by the machine protection system will be an off-energy beam. The second type of errant beam arises from the front-end system, such as a lower-than-nominal current beam pulse, a partial beam pulse or another abnormal beam arising from high-voltage arcing, unstable plasma, or mis-triggered RF for plasma generation. The beam under these conditions also becomes an off-energy beam while it is accelerated, since the RF adaptive feed forward is only for the nominal beam pulse. In this case, beam loss monitors are the first indicators of the event. The machine protection system delay is the same as for the RF truncation case. The original requirement of the machine protection system delay at the SNS is 30 µs.

The SNS machine protection system current is providing 'beam abortion' in 10–25 µs for these errant conditions.

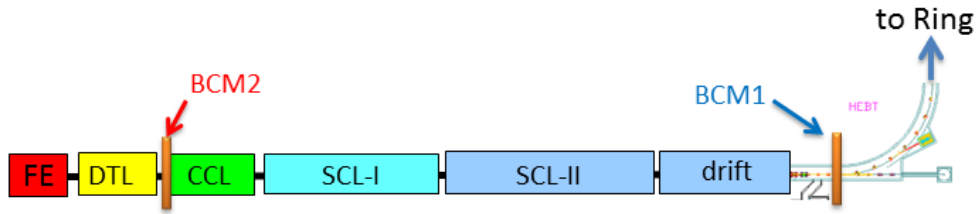

**Fig. 17:** Layout of SNS LINAC and positions of beam current monitors for errant beam monitoring. BCM, beam charge monitor; CCL, coupled cavity linac; DTL, drift tube linacs; FE, front end; HEBT, high-energy beam transport; SCL, superconducting linac.

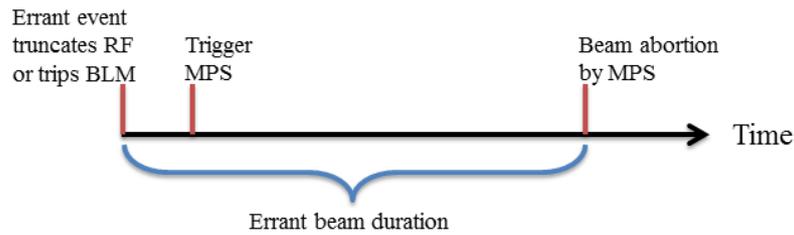

**Fig. 18:** Machine protection system delay time until abortion of beam. BLM, beam loss monitors; MPS, machine protection system.

Figure 19 shows an example of signals from beam current monitors 1 and 2 at an errant beam condition caused by RF truncation in one of the drift tube linac structures. In this example, the whole beam was lost for about 25 µs in the superconducting linac. An example of errant beam events from the front end is shown in Fig. 20. The beam current waveforms after the front end and before the superconducting linac are shown. The other waveform is the cavity field waveform in one warm linac RF structure. In this example, the beam current dropped in the middle of a pulse from the front end. Notice that cavity fields of all linac RF structures do not provide a flat field, owing to the nature of the adaptive feed forward. That portion of the beam became an off-energy beam and was lost in the linac. As mentioned earlier, beam loss monitors trigger the machine protection system and the machine protection system shuts the beam off.

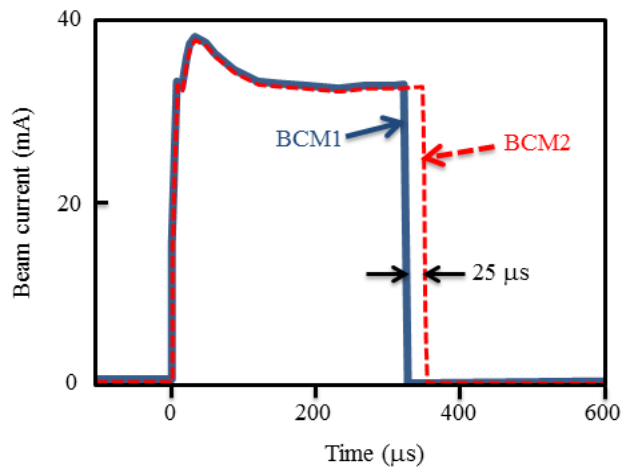

**Fig. 19:** Examples of beam current monitor (BCM) signals in the event of RF truncation in a warm linac structure

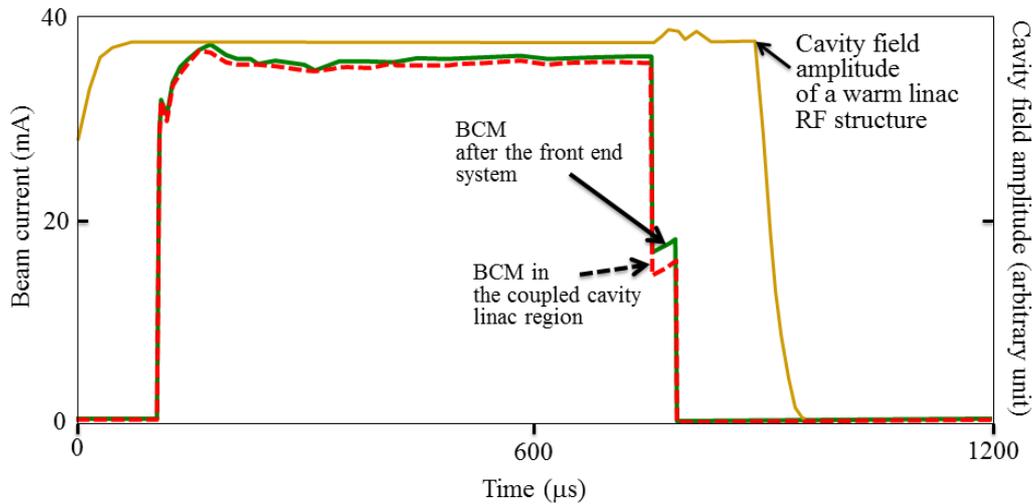

**Fig. 20:** Example of errant beam generated in the front-end system. Beam current monitor signals after the front end and in the coupled cavity linac region, and cavity field amplitude of a warm linac RF structure.

The severity of the events depends on the locations where errant beams are lost and the frequency of events that are associated with the conditioning status of errant beam sources (warm linac and front end). Usually, in the first few weeks after a long maintenance shut-down, errant beam events are more frequent. The average number of errant beam events in the past at SNS was about 30–40 times a day out of five million pulses a day. Most errant beam events result in beam loss monitor trips or sometimes small vacuum excursions. However, when similar events recur, there is a chance that an errant beam might evaporate gases and a following RF interaction could create an environment for severe discharge or arcing. The energy of one mini pulse (about 1 µs beam) at the SNS is about 24 J. Since the beam loading in the SNS superconducting RF cavities is high, the available RF power is large enough to create a dangerous discharge. Unwanted consequences from errant beam events might be additional gas or particulate contamination, surface damage that leads to RF performance degradation, and component damage, such as ceramic windows and ceramic feed-through failures. The window crack shown in the left side of Fig. 8 was caused by arcing from errant beam events. Various suggestions have been put forward to minimize the number of errant beam events, such as careful conditioning for the front end and warm linac, routine maintenance of vacuum systems, and continuous adjustment of operating parameters for warm linac structures. Presently, the frequency of errant beam events is about 10–15 times per day. In addition, a new dedicated protection system to abort the beam within 6 µs has been developed and will be used for operation in the near future [14].

## 4    Summary

Breakdown conditions as indicated in this section depend on various factors, which affect the performance of a system. For example, apparently random failure events and large scattering performances, which are commonly encountered even in the same types of system, result from the irregularity of physical and chemical properties. Therefore, it is important to take into account all possible factors in an organized way. Some factors leading to breakdown are definitely determined during design and fabrication, for example, material surface finish, process history, baking, shape, electromagnetic field configuration, pulse length, and RF frequency. Other factors are due to changes during operation, such as residual gas species, partial pressure, contaminants, radiation, circuit characteristic, temperature, history of operation, and beam conditions. Because breakdown is affected by combinations of these factors, there is no unique mechanism for a specific breakdown event.

Surface properties strongly depend on the previous operation history. It is usually found that breakdown conditions in a vacuum can be substantially improved by allowing repeated mild breakdowns or the passage of appreciable pre-breakdown current. This behaviour is called conditioning, and is a necessity in most high-power RF systems. This conditioning process should be developed in a controlled way in conformity with experimentally observed phenomena. Conditioning changes surface conditions in similar ways to breakdown mechanisms; sputtering of surface protrusions, release and relocation of trapped or condensed gases by electron bombardments, etc.

Most high-power RF systems operate with considerable energy or power, to create discharge. In many cases there is no clear boundary between conditioning and irreversible damage processes since threatening breakdown can develop quickly. If the damage is large enough, it will reduce the breakdown field, leading to high pre-breakdown current, loss of vacuum, mechanical damage, and, eventually, catastrophic failure. The amplitude of each discharge during the conditioning process can be limited by proper interlocks, such as vacuums, arc detectors, X-ray detectors, or forward RF power thresholds.

For large-scale accelerators, RF equipment protection now involves complex systems that consist of slow interlocks, fast interlocks for beam abortion, sequences to verify equipment status ready for beam operation, and logic circuits to resume normal operation automatically. The RF protection system is fully integrated with other machine protection systems, global timing systems, diagnostics systems, and control systems in modern accelerators. Fault condition input to the main machine protection system to abort beam operation should be classified based on a good understanding of the causes and consequences of a potential failure mechanism.